\documentclass[conference]{IEEEtran}

\usepackage{todonotes}
\usepackage{xcolor}
\usepackage{hyperref}
\usepackage{caption}
\usepackage{subcaption}
\usepackage{textcomp}

\color{black}

\begin{document}
\title{Automated Non-Destructive Inspection of Fused Filament Fabrication Components using Thermographic Signal Reconstruction}

\author{\IEEEauthorblockN{Joshua E. Siegel}
\IEEEauthorblockA{Computer Science and Engineering\\
Michigan State University\\
East Lansing, MI 48824\\
Corresponding author: jsiegel@msu.edu}
\and
\IEEEauthorblockN{Maria F. Beemer\\ and Steven M. Shepard}
\IEEEauthorblockA{Thermal Wave Imaging, Inc. \\ 845 Livernois St, Ferndale, MI 48220}}

\maketitle

\begin{abstract}
Manufacturers struggle to produce low-cost, robust and complex components at manufacturing lot-size one. Additive processes like Fused Filament Fabrication (FFF) inexpensively produce complex geometries, but defects limit viability in critical applications. We present an approach to high-accuracy, high-throughput and low-cost automated non-destructive testing (NDT) for FFF interlayer delamination using Flash Thermography (FT) data processed with Thermographic Signal Reconstruction (TSR) and Artificial Intelligence (AI). A Deep Neural Network (DNN) attains $95.4$\% per-pixel accuracy when differentiating four delamination thicknesses $5$mm subsurface in PolyLactic Acid (PLA) widgets, and $98.6$\% accuracy in differentiating acceptable from unacceptable condition for the same components. Automated inspection enables time- and cost-efficient $100$\% inspection for delamination defects, supporting FFF's use in critical and small-batch applications. 
\end{abstract}

\IEEEpeerreviewmaketitle
\section{Motivation}
Modern manufacturing challenges assumptions about how quickly something can be produced and through which processes. Industry $4.0$ is evolving toward unit-size customization, while innovation accelerates in pursuit of reduced cost, improved efficiency, and enhanced quality. Today, rapid prototyping, rapid tooling, late differentiation and the ability to reliably produce complex geometries are prerequisites for success. 

Fused Filament Fabrication (FFF) is a form of Additive Manufacturing (AM) addressing these technical and economic needs. In FFF, polymer feedstock, typically in the form of thermoplastic pellets or filaments, is melted and extruded into geometries that fuse and solidify to form complex geometries layer-by-layer. FFF machines are low-cost, geometrically flexible, capable of holding tight tolerances and using varied materials, and have large work envelopes. Recent advances increase production rate, further improving FFF's fitness for custom manufacturing-as-a-service\cite{LVC_paper}. Such high-quality, granular customization is necessary in healthcare, where AM is used to create surgical implants or assistive devices\cite{huang2013additive} or in aerospace, where strong, lightweight components are essential. 

Despite FFF's potential, extrusion physics pose challenges. Imprecise machine positioning, environmental changes, feedstock variability and nozzle clogs can lead to interlayer delamination, voids, inclusions, subsurface defects, and non-colinear deposition. Fault modalities might also be malicious in nature; for example infill path, layer thickness, or fan-speed modification \cite{Gao:2018:WSY:3279953.3264918} could be modified in G-code or in object geometry files, or introduced via firmware-resident command-substitution extruder rate attacks\cite{implications_malicious}. Defective components resulting from either process, or from design deficiencies, may harbor catastrophic weaknesses\cite{10110813552540210441166,206156}. 

Destructive testing identifies such faults at significant time and economic cost, while also requiring the construction of destructible widgets. Process variability additionally means tested samples may not be representative. Non-destructive inspection like radiography addresses these challenges but creates time and expense bottlenecks.

Automated, low-cost, high-throughput and reliable Non-Destructive Testing (NDT) with $100$\% inspection would allow FFF's use in applications from consumer products to mission-critical systems. This paper presents a concept for non-destructive evaluation of components' internal geometries using flash thermograph (FT) techniques in conjunction with Artificial Intelligence (AI).

In Section~\ref{prior_art}, we review contemporary uses for NDT as applied to FFF and explore thermography as an NDT method, using TSR processing of acquired data sequences for identification of FFF delamination defects. We develop representative defect models in Section~\ref{simulated_results} and prove the viability of automated TSR feature classification for different data sequences. In Section~\ref{real_data}, we describe collecting real-world TSR data from samples of delaminated components imaged using flash thermography and TSR processing, and we subsequently train a deep neural network in Section~\ref{dnn_design} to differentiate between normal and abnormal TSR features. Finally, we present our results and consider future improvements in Sections~\ref{results} and~\ref{conclusion}.

\section{Prior Art and Hypothesis}\label{prior_art}
FFF failure modes include voids (air pockets), inclusions (foreign material deposits) and poor interlayer adhesion (kissing bonds, where there is mechanical contact between surfaces but no adhesion, or delamination, a case where adhesion between layers is only partial). In all cases, there exists a region within the component with different thermal properties than the surrounding materials. 

Seppala and Migler use passive infrared imaging to detect delamination in-process by monitoring the extrudate relative to its glass transition temperature\cite{seppala2016infrared}, identifying early-solidification but potentially missing defect sources such as feedstock starvation, extruder occlusion or clogging, and imprecise axial travel.

Post-print inspection techniques include ``active thermography,'' wherein an object is heated using a high-energy, plasma discharge flashlamp and imaged using an infrared camera to measure the surface temperature as the object cools to a quasi-equilibrium state. In flash thermography, the most widely used active NDT configuration, the excitation is a brief ($<5$ msec) pulse of light from xenon flash lamps that is applied uniformly to the part surface, effectively creating $1$-dimensional thermal diffusion from the surface into the sample volume\cite{anst_handbook}. The presence of subsurface anomalies (e.g. voids, delamination, inclusions, layer interfaces or internal structure) disrupt the local cooling process, causing the pixel surface temperature vs. time history to deviate from its normal (unobstructed one-dimensional cooling) response. The transient response of each pixel encodes information about material properties and component geometry, including subsurface defects. 

The emergence of thermography as an NDT method was enabled by the invention and commercial availability of the IR camera in the late 1960's and 1970's\cite{bjork1967aga,carlomagno1976unsteady}. Early efforts were limited to detection of severe defects with large aspect (diameter to depth) ratio that could be identified visually in the post-excitation image sequence as local hot or cold spots relative to a defect-free background\cite{milne1985non,thompson1990thermographic}. However, as the sensitivity, speed and resolution of camera technology improved, methods for through-pixel analysis emerged\cite{maldague1996pulse,rajic2002principal,shepard2003reconstruction,vavilov2010thermal} to provide temporal noise reduction and enhance each pixel time history, enabling quantitative analysis, automation and detection of smaller and subtler features, and enabling material characterization, in addition to flaw detection. 

Maldague introduced the ``rule of thumb'' detectable condition that the aspect ratio of a defect should be greater than $~2$.\cite{maldague2001theory} Almond et al. utilize active thermography and show that trapped heat flows laterally around defects, creating inclusion-size dependent temperature contrast\cite{almond1994defect}. This work considers spatially discrete, large-diameter defects but does not characterize extended interface where no edges or shape cues are present, which may be significant in strength-critical applications. More recently, Oswald-Tranta et al. showed the same for additively-manufactured parts\cite{oswald2016comparison}. This work similarly considers large-diameter faults, but does not characterize smaller voids or layer delamination which may be significant for some applications. 

Shepard et al. introduced Thermographic Signal Reconstruction (TSR) (2003)\cite{shepard2003reconstruction} which provides improved detection of low contrast and low aspect ratio features and the ability to detect and measure the characteristics of extended interfaces. \cite{twi_qirt} TSR provides noise reduction by fitting the logarithmic temperature vs. time history with a low order polynomial, and signal enhancement by taking the 1\textsuperscript{st} and 2\textsuperscript{nd} derivatives (with respect to log time) of the noise-reduced replica. \cite{shepard2003temporal,balageas2012}

For active thermography, the resulting derivative signals closely follow results from analytical and numerical modeling\cite{twi_qirt} and illustrate variations in the thermophysical properties of the instrumented object, precisely indicating intentional and accidental material changes or manufacturing defects. Shepard and Beemer describe a the dependence of extended interface detection on the ratio of layer thermal effusivities\cite{shepard2015advances}, and add the additional dependence for aspect ratio for discrete defects\cite{twi_qirt}. Images of the derivatives at a particular time provide a significantly more detailed view of subsurface features compared to the unprocessed IR camera output, reducing or eliminating artifacts due to reflection or nonuniform excitation\cite{shepard2007flash}. Additional images, based on signal attributes (e.g. derivative extrema or threshold crossings) enable measurement of thermophysical properties, automatic defect recognition and quantitative part-to-part comparison\cite{shepard2007automated,shepard_fingerprints}.

While thermography has been applied to FFF, TSR has not been widely studied in this application. Despite commonly-used FFF materials' (Acrylonitrile Butadiene Styrene [ABS] or PolyLactic Acid [PLA]) poor thermal conductivity, TSR may yield features useful for identifying sub-surface defects. These signals may be difficult for humans to differentiate when manually examining data and pixel-wise classification would be infeasible for whole component inspection, but these features may be readily differentiable using Deep Learning, facilitating rapid and comprehensive component inspection capable of improving production quality. 

We propose using Artificial Intelligence (AI) in the form of a Deep Neural Network (DNN) to examine TSR features and automatically segment pixel regions as being normal or abnormal, and if an abnormal delamination, the degree to which the region is not full adhered. We select Deep Learning instead of techniques like SVM because Deep Learning deals well with potential nonlinearities and better handles the high volume data comprising individual pixel time-series. Successful AI-backed TSR (AI-TSR) would increase inspection throughput and performance without increasing labor cost, supporting increased production rates and 100\% inspection of every production run. Such improvements would increase FFF's applicability in critical applications and generally improve the quality of AM components. The following sections consider the development of an artificial neural network designed to classify per-pixel TSR feature vectors. 

\section{Proof of Concept (Synthetic Data)}\label{simulated_results}
To prove TSR's applicability to FFF components, we developed representative hard-to-differentiate synthetic cooling profiles, fit TSR features, and trained a DNN to differentiate pixel features into two classes representing ``normal'' and ``delaminated'' regions. 

\begin{figure*}
\centering
\begin{subfigure}{.5\textwidth}
 \centering
 \includegraphics[width=.8\linewidth]{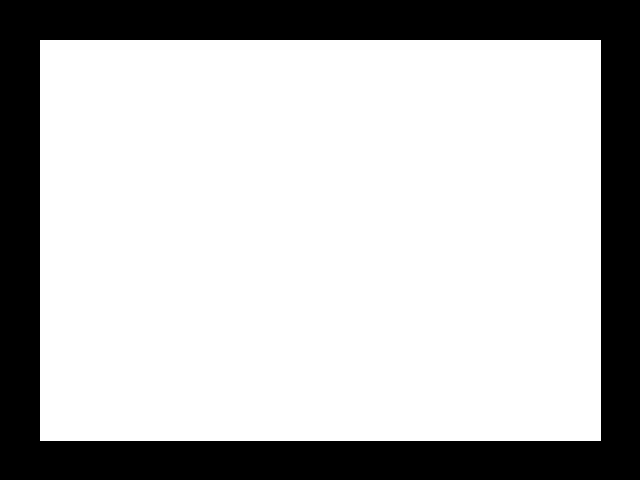}
 \caption{This image shows the ground truth test video segmentation. }
 \label{fig:sub1}
\end{subfigure}%
\begin{subfigure}{.5\textwidth}
 \centering
 \includegraphics[width=.8\linewidth]{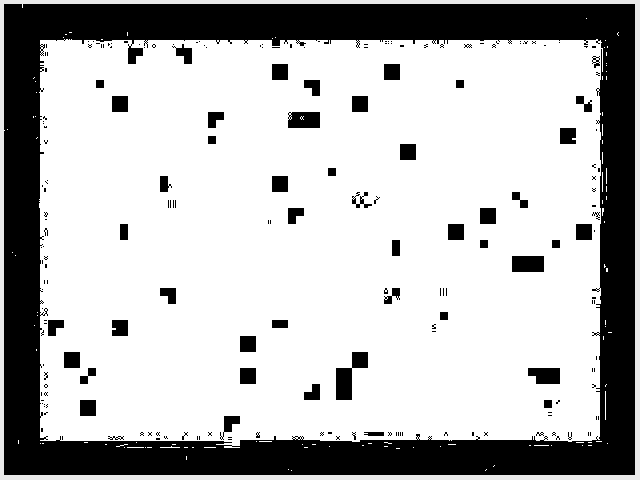}
 \caption{This image shows the classifier's predicted labels from noisy data. }
 \label{fig:sub2}
\end{subfigure}
\caption{These figures compare the true and classified states of the reference video. }
\label{sample_video_vs_classifier_output}
\end{figure*}

We first developed software to create synthetic thermographic video files taking as input region boundaries, temperature vs. time profiles, and pixel noise and range (RGB/HSV) parameters. We defined two temperature-time profiles (normal and delaminated) approximated from known material behavior using 8\textsuperscript{th}-order polynomials and constrained R, G and B values to $[0-254]$ to simulate pixel saturation, with each pixel having a random, low-amplitude Gaussian noise parameter to corrupt the signal and emulate the camera sensor's output. From these inputs and constraints, we generated three $640$x$480$ pixel videos: one training video represented the ``normal'' class, another training video represented the ``delaminated'' class, and a third testing video comprising both classes, with one series as a centralized rectangle and the other series populating the border (see Figure~\ref{fig:sub1}). Randomized pixel noise ensured that the composite video featuring both classes represented true outsample data with no pixels duplicated from the ``pure'' training videos.

DNN input features were created by loaded the two training videos ($100$\% normal and $100$\% delaminated) and fitting each pixel's log-log time-series to an $8$th-order polynomial using least-squares, to provide an effective low-pass filter\cite{shepard2003reconstruction}. After fitting, we computed each pixel polynomial's first and second derivatives and concatenated each coefficient into $21$-element feature vectors which were then associated with a class label.

The result was a series of $640$x$480$=$307,200$ feature vectors for each class. We split these feature vectors into training ($80$\%) and testing ($20$\%) sets at random and without repetition. These data were fed into a DNN with three layers possessing $16$, $32$ and $16$ fully connected units, similar to the DNN described in \cite{afci}. Our learning rate was initially $1$e$-7$, with a decay step of $1,000$ and a decay rate of $0.9$. We selected a batch size of $512$ pixel time series to balance overfitting with generalizability, and we allowed for up to $100,000$ steps but curtailed learning with early stopping should the validation losses increase three consecutive times $100$ steps apart. The model was implemented in TensorFlow. 

This method obtained $95.7$\% in-sample accuracy. Validating on TSR features extracted from the pure outsample video, we obtained an accuracy of $92.8$\%. The classifier's output labels and ground truth are compared visually in Figure~\ref{sample_video_vs_classifier_output}. 

These early results were promising given the subtle difference in the representative polynomials. Next, we attempted classification on real-world data captured from FFF components. 

\section{Real-World Data}\label{real_data}
With the concept of automated TSR classification using AI validated for synthetic data, we continued on to image samples of typical FFF defects. We began by manufacturing rectangular prisms with various delamination states, ranging from fully-fused adhesion to kissing bonds, in which layers boundaries are mechanically in contact but almost entirely unfused. 

\subsection{Producing Sample Components}
Delamination defects occur primarily in widgets produced by layer-based material deposition techniques such as FFF. The FFF approach utilizes one or more heated nozzles to extrude thermoplastic material feedstock into thin filaments, precisely depositing the resulting molten polymer such that the heat of the extrusion causes the polymer to locally exceed its glass transition temperature. These filaments are initially deposited onto a work surface (the AM machine bed), and subsequently onto existing, deposited filament. The high temperature of the extrudate ultimately causes intra-layer fusion upon cooling. By repeating this process layer-by-layer, a strong, solid object is produced -- assuming the process and feedstock are well-controlled. In other cases, the extruded filament may fail to fuse to supporting layers as a result of having too large a gap between the extruded filament and the underlying support structure, due to variation in temperature of the support structure, nozzle, or filament, due to variations in feedstock quality, or other reasons. 

To emulate delamination as might be encountered in a typical FFF process, samples were produced using a Dremel 3D20 FFF AM machine, selected for its material handling capabilities and positioning repeatability as well as the ability for the controller to read G-Code directly from files stored to the AM machine's memory card. The 3D20 is capable of processing Acrylonitrile Butadiene Styrene (ABS) and Polylactic Acid (PLA) feedstock, though the machine is better-suited to PLA production due to the lack of a heated bed (a closing door and removable lid do help to maintain a consistent internal temperature inside the machine's build chamber). The ability to process G-code from a file simplified manual nozzle position modification, enabling delamination to be precisely induced at a particular depth and with a specific thickness. 

\begin{figure}[ht]
\centering
 \includegraphics[width=0.35\textwidth]{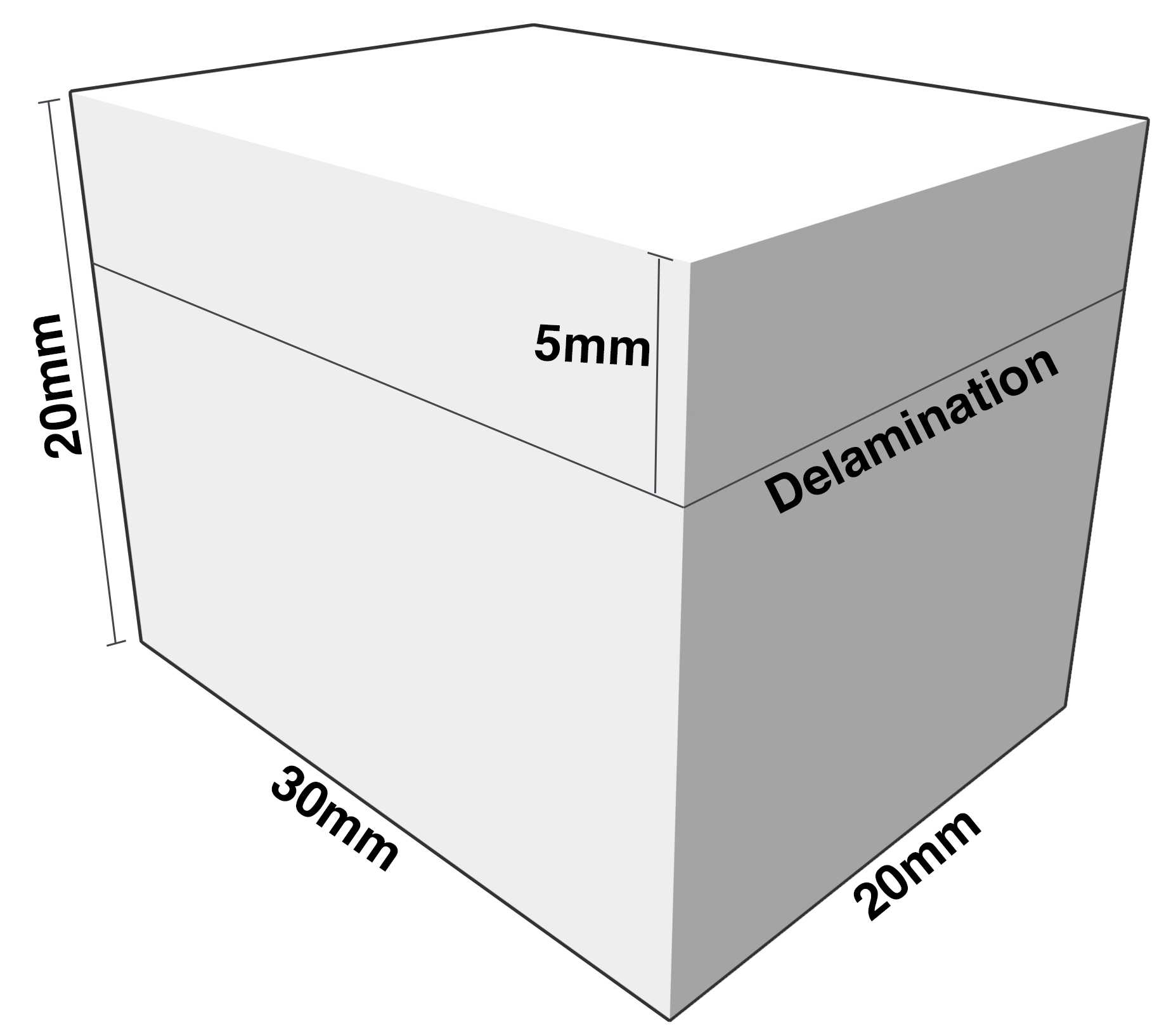}
 \caption{This figure shows the FFF geometry, of the rectangular prism, with a delamination $5$mm from the top surface. }
 \label{delam_drawing}
\end{figure}

The ``normal'' reference piece is $20$w x $30$d x $20$h mm with solid ($100$\%) infill, dimensions selected as being thick enough to prevent back-wall effects from impacting thermal diffusion for the given simulated delaminations. We additionally printed three ``abnormal'' variations, manually modifying the slicer's output G-Code to insert a delamination of $0.1$, $0.2$ and $0.3$mm located $5$mm below the top surface of the prism. The delaminations represent $33$\%, $66$\% and $100$\% of the nominal interlayer height (in practice, the gaps are slightly smaller due to extrudate swelling and sagging). A representative figure showing the (delaminated) prism geometry is shown in Figure~\ref{delam_drawing}.

The samples were printed using a new roll of black $1.75$mm ``Solutech'' PLA feedstock to minimize the impact of water absorption on deposition solidity and to improve the component's emissivity and therefore thermal imageability. The machine used the settings in Table~\ref{print_settings}.

\begin{table}[]
\centering
\caption{This table shows the settings used to generate G-Code to print representative rectangular prismatic samples on a Dremel 3D20 AM machine (3D printer). Additional Z-axis travel was added within the G-code to induce delamination $5$mm below the component's surface. }
\label{print_settings}
\begin{tabular}{rc}
Machine Settings & \multicolumn{1}{l}{} \\ \hline
\multicolumn{1}{|r|}{\textbf{Make/Model}} & \multicolumn{1}{c|}{\textit{Dremel 3D20}} \\ \hline
\multicolumn{1}{|r|}{\textbf{Print Material}} & \multicolumn{1}{c|}{\textit{PLA}} \\ \hline
\multicolumn{1}{|r|}{\textbf{Extrusion Temperature}} & \multicolumn{1}{c|}{\textit{220C}} \\ \hline
\multicolumn{1}{|r|}{\textbf{Heated Build Plate}} & \multicolumn{1}{c|}{\textit{No}} \\ \hline
\multicolumn{1}{|r|}{\textbf{Print Speed}} & \multicolumn{1}{c|}{\textit{120mm/s}} \\ \hline
\multicolumn{1}{|r|}{\textbf{Traversal Speed}} & \multicolumn{1}{c|}{\textit{120mm/s}} \\ \hline
\multicolumn{1}{|r|}{\textbf{Infill}} & \multicolumn{1}{c|}{\textit{100\%}} \\ \hline
\multicolumn{1}{|r|}{\textbf{Infill Pattern}} & \multicolumn{1}{c|}{\textit{Rectilinear}} \\ \hline
\multicolumn{1}{|r|}{\textbf{Layer Height (First)}} & \multicolumn{1}{c|}{\textit{0.4mm}} \\ \hline
\multicolumn{1}{|r|}{\textbf{Layer Height (Subsequent)}} & \multicolumn{1}{c|}{\textit{0.3mm}} \\ \hline
\multicolumn{1}{|r|}{\textbf{Surface Shells}} & \multicolumn{1}{c|}{\textit{2}} \\ \hline
\multicolumn{1}{|r|}{\textbf{Support Structure}} & \multicolumn{1}{c|}{\textit{None}} \\ \hline
\end{tabular}
\end{table}

While the produced prisms had visual imperfections on the top layers, they were deemed adequate as representative samples as the defects were on non-structural layers. The internal layers were printed completely and fully adhered to one another. 

\subsubsection{Sample Imaging}
\begin{figure}[ht]
\centering
 \includegraphics[width=0.49\textwidth]{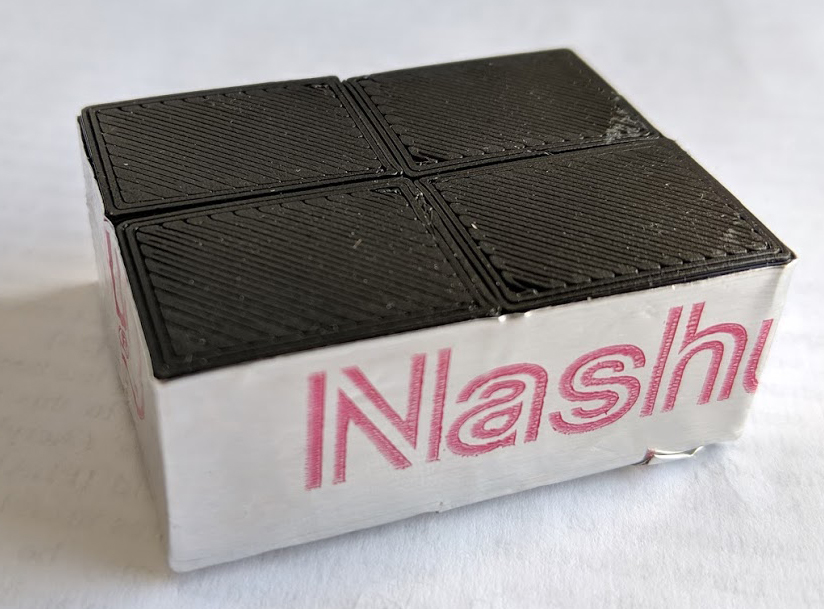}
 \caption{The four FFF samples were taped together using a metallized tape. The tape and close proximity of the samples minimized convection effects near the sample edges. }
 \label{samples_taped}
\end{figure}

\begin{figure}[ht]
\centering
 \includegraphics[width=0.49\textwidth]{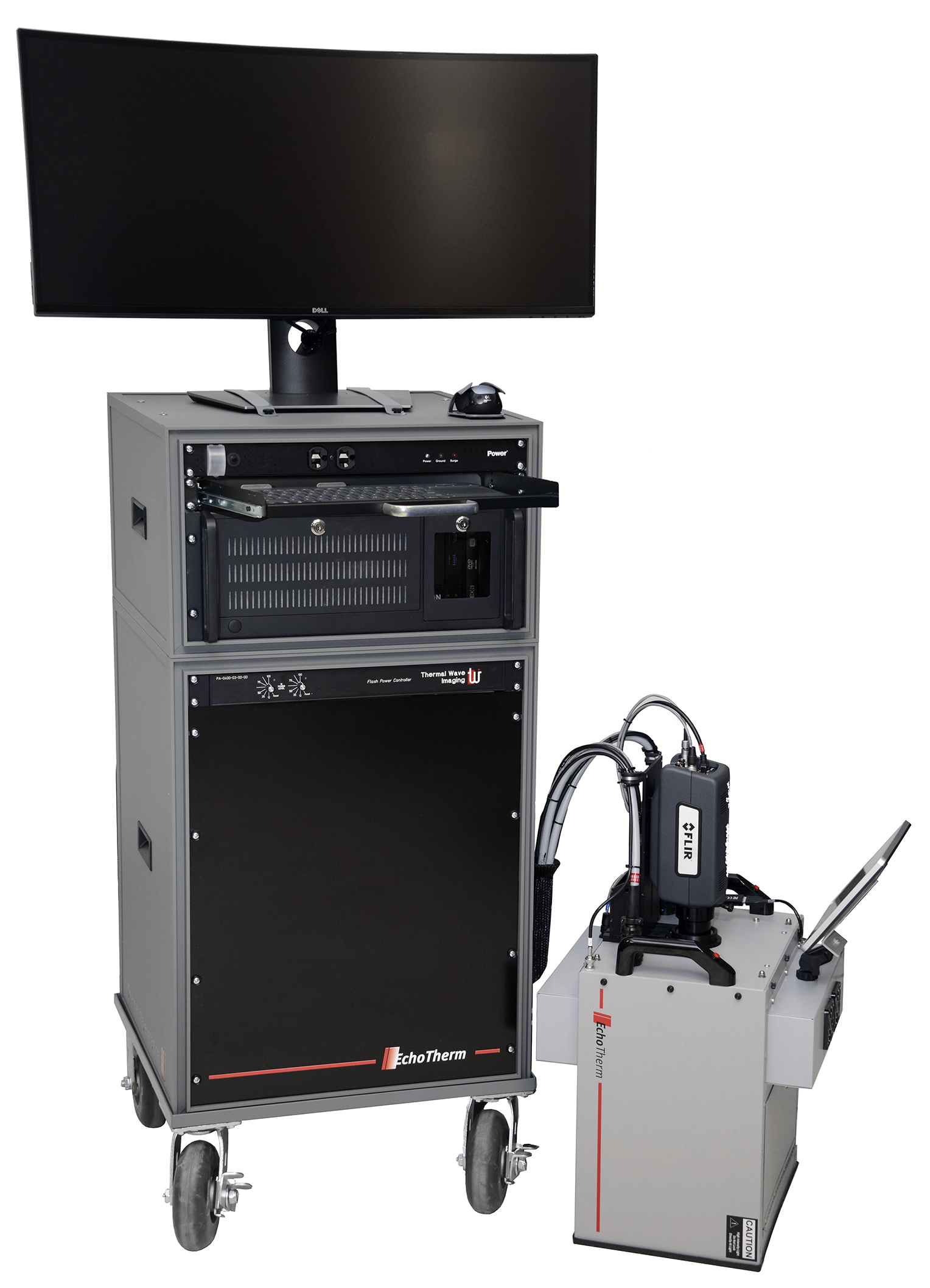}
 \caption{EchoTherm\textregistered~is an active thermography system used to heat components and subsequently capture cool-down data, information informing the creation of TSR features used as input for AI-backed condition grading and classification. }
 \label{echotherm}
\end{figure}

\begin{figure}[ht]
\centering
 \includegraphics[width=0.49\textwidth]{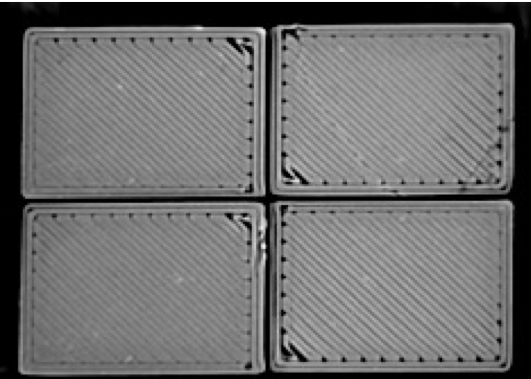}
 \caption{This figure shows the color image of the four cubes taped together. Clockwise from upper right is the baseline (normal) cube, $0.1$mm delamination, $0.3$mm delamination, and $0.2$mm delamination. All delaminations are $5$mm from the surface. }
 \label{color_image}
\end{figure}

\begin{figure}[ht]
\centering
 \includegraphics[width=0.49\textwidth]{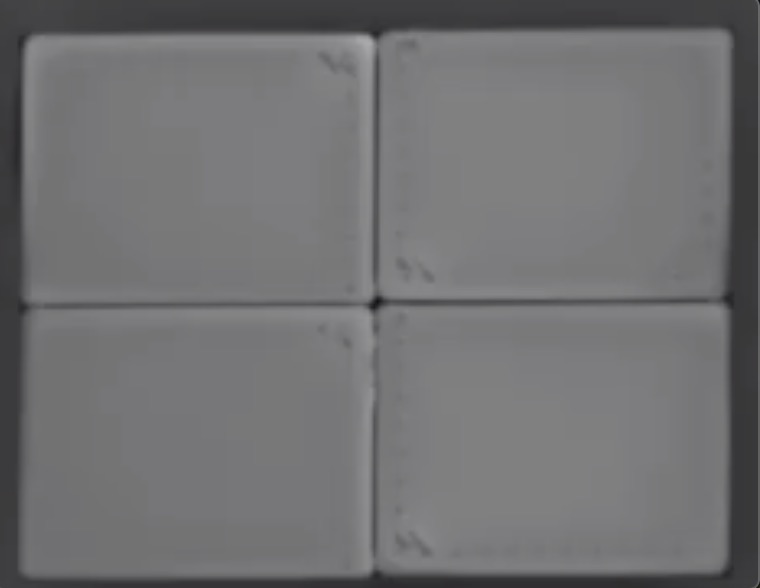}
 \caption{This figure shows a sample thermal image of the four cubes taped together immediately after heating. }
 \label{sample_thermal_image}
\end{figure}

Before imaging, the samples were positioned in contact with one another and held together using metallized tape wrapped around the perimeter to minimize edge convection effects (Figure~\ref{samples_taped}). They were then placed on an optical table and simultaneously thermally excited using a high-energy plasma flashlamp in the form of Thermal Wave Imaging, Inc's EchoTherm\textregistered~system (Figure~\ref{echotherm})\footnote{EchoTherm\textregistered, Thermal Wave Imaging, Inc., \url{www.thermalwave.com}.}. During this process, the series of cubes was imaged from the top-down (Figure~\ref{color_image} shows the RGB image, and Figure~\ref{sample_thermal_image} shows the initial thermal image of the cubes undergoing cooldown). The thermal transient response was recorded for $240$ seconds with a Flir A6751sc camera recording $640x512$px, later cropped to $236$x$182$px, at $15$ frames per second (plastic's low dispersion rate allows the use of slower, less costly cameras than those required to generate TSR features for most composites or alloys). Based on the number of samples imaged simultaneously and the field of view, this yielded $>=5,000$ pixel time series per defect class. The long cooldown allowed the heat to fully soak, and PLA's poor conductivity allowed acquisition to terminate before the injected heat reached the backwall of the sample. Data were captured to a video file and exported using Thermal Wave Imaging, Inc.'s Virtuoso\textregistered~\footnote{Virtuoso\textregistered, Thermal Wave Imaging, Inc., \url{www.thermalwave.com}.} to a series of CSV files for post-processing. 

\section{TSR and Neural Network Implementation}\label{dnn_design}
After capturing thermal imaging data from real FFF samples using commercially-available hardware and software, we sought to automate TSR feature differentiation. 

\begin{figure*}[ht]
\centering
 \includegraphics[width=1.0\textwidth]{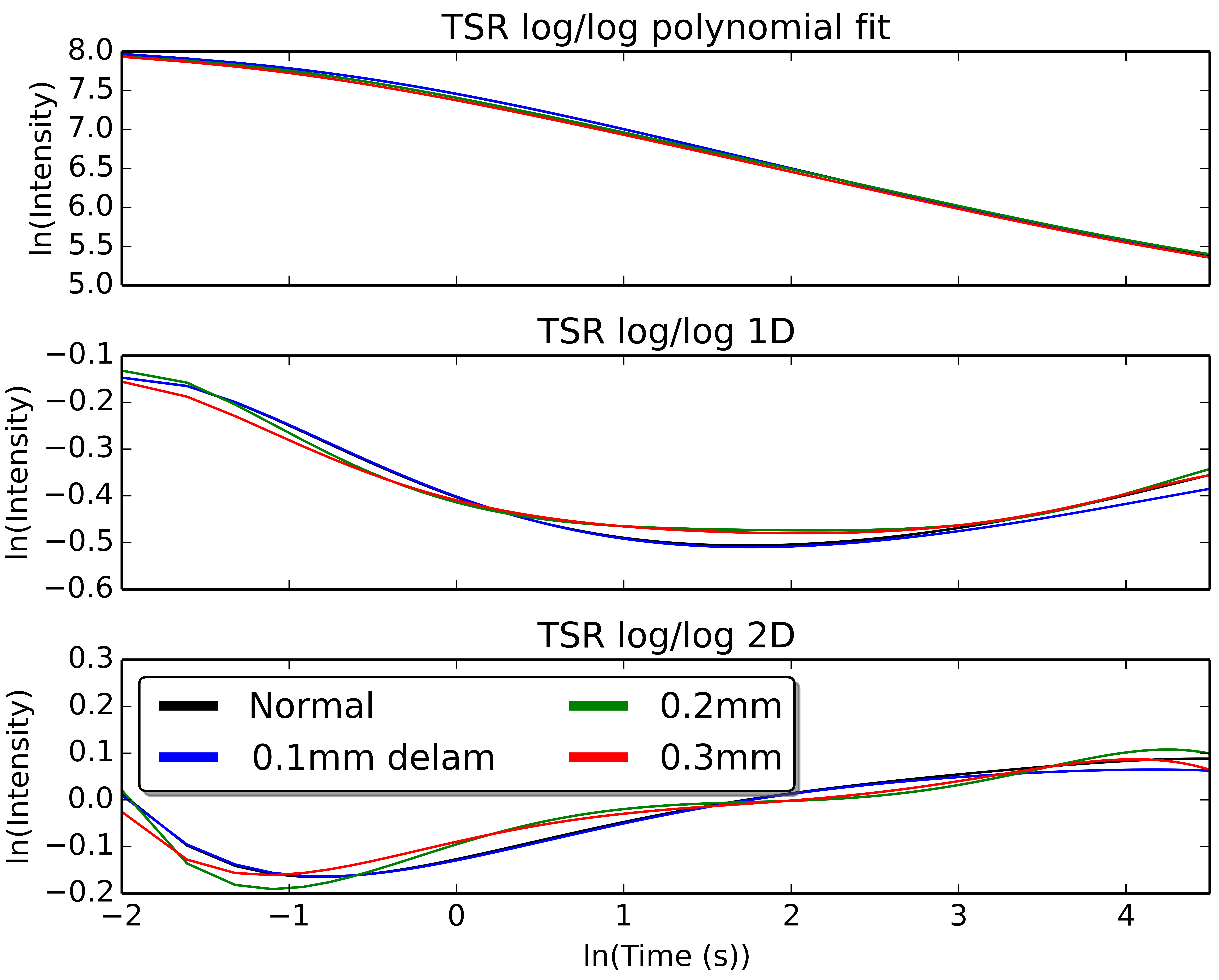}
 \caption{Plots compare log-log temperature versus time profiles and derivatives for four delamination cases (normal through $0.3$mm delamination). Though difficult to discern visually, the coefficients describing these polynomials create feature vectors useful for robust classification.}
 \label{tsr_representative}
\end{figure*}

From the four recorded pixel time series ranging from full adhesion to $0.3$mm delamination, we initially fit $8$th-order polynomials to the log-log time-temperature series captured by the video file. Pixels regions were segmented manually, using the visible-spectrum image of the prisms to define boundary locations between classes. A small (~$5$px) boundary was discarded on all sides of the region to minimize model fit to pixels possessing conductivity- and conduction-related edge effects. In Figure~\ref{tsr_representative}, we see that the behavior of the log-log fitted TSR for representative pixels for each of the four states (normal and three varied delaminations) appears similar. However, the first and second derivatives vary widely. In particular, the tail behavior of the second derivatives diverges across the classes, with similarities between normal and $0.1$mm delamination and between the $0.2$ and $0.3$mm delaminations. This indicates that the largest separation between classes occurs when the delamination thickness exceeds half of the nominal layer height ($0.3mm/2=0.15mm$), which is critical from a strength perspective.

We initially tested these $8$th-order polynomial coefficients using the same DNN as that designed to classify the synthetic data. However, the real data differed from the synthetic data in several ways, including the frequent appearance of negative coefficients, which would not work well with the synthetic DNN model's activation functions ($ReLU$). Additionally, real data had more variation across pixels than synthetic data, both due to noise and as a result of conduction effects. As a result, performance when reusing the initial DNN was poor -- simply running the earlier network resulted in a degenerate solution where the classifier guessed the same class for each input vector.

We needed to simultaneously optimize the TSR fit parameters as well as the DNN design. Using a fourth-order polynomial provided a balance between the closeness of TSR feature fit, potential for overfit, and the resulting size of the feature vector and computational complexity of the DNN. The TSR fit was performed in Virtuoso\textregistered~software using a 4\textsuperscript{th} order polynomial, and skipping the earliest frames, which were saturated due to the limited dynamic range of the camera calibrated for room temperature operation.

The resulting feature set included the polynomial coefficients of the fit of the log-log data and of its first and second derivatives, for a total of $15$ features per pixel. To further improve classification performance, a scaling step was added wherein each element of the set of training feature vectors was scaled to have a mean of zero and a standard deviation of one. Scaling factors were determined using only the training set as input to avoid contamination of the model with the use of testing data, and the same parameters were used to re-scale testing and validation data. 

For developing the new DNN model, data were split $80$/$20$\% training and testing, respectively, and $10$\% of the training set was kept out as a validation set. To improve performance for unseen cases, we conducted data augmentation by modifying the feature vectors with randomized noise on each feature allowing up to $5$\% deviation from each coefficient's raw value. We generated $50$ synthetic sets and maintained the features from the baseline set for use in the final training set. 

\begin{figure}[ht]
\centering
 \includegraphics[width=0.3\textwidth]{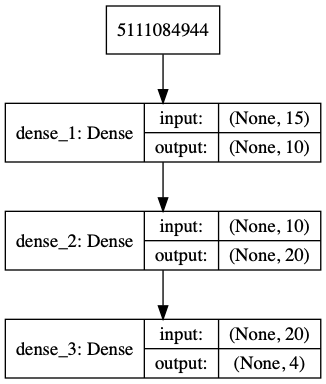}
 \caption{The DNN designed is a three layer model consisting entirely of dense layers using the $tanh$ activation function. $5,111,084,944$ is the number of input pixel time series used for training, each with $15$ associated features. The four states reflect a one-hot vector for normal output or one of three degrees of delamination defect.} 
 \label{model_design}
\end{figure}

A new DNN was constructed based upon the earlier model, but accounting for changes in the type and distribution of data in the new set. The new model had three layers, an input layer with $10$ neurons and a $tanh$ activation function to account for negative values, a hidden layer with $20$ neurons and a $tanh$ activation function, and an output layer with four neurons (supporting one-hot state encoding) and a $softmax$ activation function. The network is visualized in Figure~\ref{model_design}. The learning rate was $1e-5$, batch size $2,048$, and the network used the Adam optimizer. Performance metrics included test accuracy, with a categorical cross entropy loss function. There was no early stopping or decay for the learning rate, and the model was implemented in TensorFlow using the Keras API. While this model demonstrated fast computation, quick convergence and robust performance, others may work similarly well. 

\begin{figure}[ht]
\centering
 \includegraphics[width=0.49\textwidth]{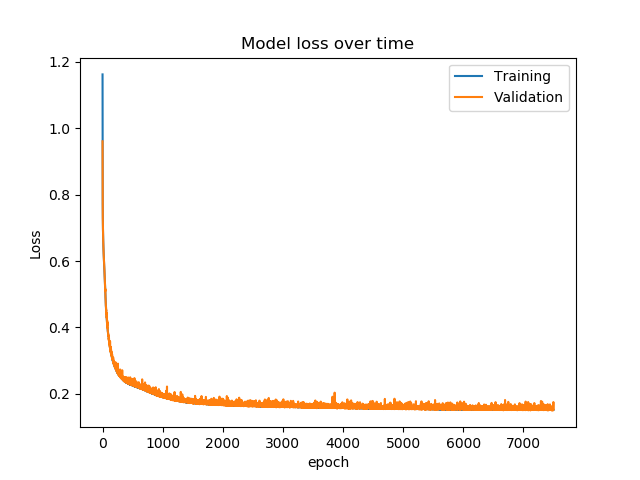}
 \caption{Both training and validation loss increased consistently over time. Training was stopped when the validation loss appeared to level out, and before it began to increase, indicating potential overfit.}
 \label{val_loss}
\end{figure}

The model was trained until the validation loss leveled off and before this loss began to increase(shown in Figure~\ref{val_loss}). In practice, this was approximately $7,500$ epochs. Results appear in Section~\ref{results}. 

\begin{figure}[ht]
\centering
 \includegraphics[width=0.49\textwidth]{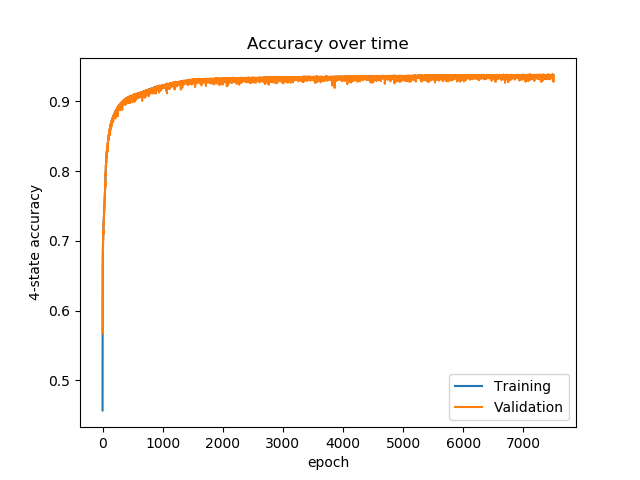}
 \caption{Training and validation set accuracy improved over time. While not strictly-monotonic, the performance continued to improve on average until training the specified number of epochs was reached. }
 \label{accuracy}
\end{figure}

\section{Results}\label{results}
Using real data, the four-state accuracy improved over time, attaining a peak of $95.4$\% on the validation set as shown in Figure~\ref{accuracy}. The resulting test set confusion matrix is shown in Table~\ref{conf_four_state}.

\begin{table*}[]
\centering
\caption{This table shows the classifier's performance when attempting to identify the precise thickness of the delamination.}
\label{conf_four_state}
\begin{tabular}{rcccc}
\cline{2-5}
\multicolumn{1}{l|}{} & \multicolumn{1}{l|}{\textit{Predicted Normal}} & \multicolumn{1}{l|}{\textit{Predicted 0.1mm}} & \multicolumn{1}{l|}{\textit{Predicted 0.2mm}} & \multicolumn{1}{l|}{\textit{Predicted 0.3mm}} \\ \hline
\multicolumn{1}{|r|}{\textit{Actually Normal}} & \multicolumn{1}{c|}{1152} & \multicolumn{1}{c|}{83} & \multicolumn{1}{c|}{1} & \multicolumn{1}{c|}{18} \\ \hline
\multicolumn{1}{|r|}{\textit{Actually 0.1mm}} & \multicolumn{1}{c|}{61} & \multicolumn{1}{c|}{1241} & \multicolumn{1}{c|}{0} & \multicolumn{1}{c|}{12} \\ \hline
\multicolumn{1}{|r|}{\textit{Actually 0.2mm}} & \multicolumn{1}{c|}{6} & \multicolumn{1}{c|}{6} & \multicolumn{1}{c|}{1377} & \multicolumn{1}{c|}{11} \\ \hline
\multicolumn{1}{|r|}{\textit{Actually 0.3mm}} & \multicolumn{1}{c|}{19} & \multicolumn{1}{c|}{14} & \multicolumn{1}{c|}{19} & \multicolumn{1}{c|}{1408} \\ \hline
\multicolumn{1}{l}{} & \multicolumn{1}{l}{} & \multicolumn{1}{l}{} & \multicolumn{1}{r}{n=} & \multicolumn{1}{l}{5429} 
\end{tabular}
\end{table*}

The four-state accuracy indicates that not only is it possible for AI to differentiate normal from abnormal FFF components using TSR features, but it is also reliable at classifying the significance of the problem in cases where some delamination (limited by gap percentage or delamination cross-sectional area) may be allowable. 

From the four-state classification results, we then generated two different two-state confusion matrices. The first, normal/abnormal, appears in Table~\ref{conf_two_state} along with performance metrics including accuracy ($96.5$\%), precision ($97.6$\%), and recall ($97.9$\%).

\begin{table}[]
\centering
\caption{This table characterizes binary classification performance for the one-versus-all case (normal vs. any size defect)}
\label{conf_two_state}
\begin{tabular}{lll}
\cline{2-3}
\multicolumn{1}{l|}{} & \multicolumn{1}{l|}{\textit{Predicted Acceptable}} & \multicolumn{1}{l|}{\textit{Predicted Unacceptable}} \\ \hline
\multicolumn{1}{|r|}{\textit{Actually Acceptable}} & \multicolumn{1}{c|}{1152} & \multicolumn{1}{c|}{102} \\ \hline
\multicolumn{1}{|r|}{\textit{Actually Unacceptable}} & \multicolumn{1}{c|}{86} & \multicolumn{1}{c|}{4088} \\ \hline
 & \multicolumn{1}{r}{\textit{n=}} & \textit{5429}
\end{tabular}
\end{table}

The second two-state derivation from the four-state classifier's results is a model where small delaminations ($< \frac{1}{2}$ $layer$ $height$) are deemed acceptable. Such a defect might be acceptable where some risk is tolerable, for example in a non-load-bearing consumer product where a failure results in a return, not a fatality. These results are shown in Table~\ref{conf_two_state_defect_ok}. 

\begin{table}[]
\centering
\caption{This table characterizes binary classification performance for the acceptable-unacceptable case (normal or small defect vs. large defect)}
\label{conf_two_state_defect_ok}
\begin{tabular}{lll}
\cline{2-3}
\multicolumn{1}{l|}{} & \multicolumn{1}{l|}{\textit{Predicted Acceptable}} & \multicolumn{1}{l|}{\textit{Predicted Unacceptable}} \\ \hline
\multicolumn{1}{|r|}{\textit{Actually Normal}} & \multicolumn{1}{c|}{2538} & \multicolumn{1}{c|}{31} \\ \hline
\multicolumn{1}{|r|}{\textit{Actually Abnormal}} & \multicolumn{1}{c|}{45} & \multicolumn{1}{c|}{2815} \\ \hline
 & \multicolumn{1}{r}{\textit{n=}} & \textit{5429}
\end{tabular}
\end{table}

In this case, performance improves. Accuracy becomes $98.6$\%, with precision ($98.9$\%), and recall ($98.4$\%). As with the earlier results, these numbers were captured from a validation set, though it is difficult to conceptualize performance in a meaningful way using confusion matrices. Visual explanations are often more meaningful. 

In order to test on representative outsample data without requiring printing and re-imaging additional components, we generated synthetic outsample data. To do so, the base coefficients for every pixel in the image were loaded to memory and then modified with random noise of $\pm3$\%. This reduced the likelihood of overfit providing artificially-high accuracy results. Each pixel was then classified using the trained model.

\begin{figure}[ht]
\centering
 \includegraphics[width=0.49\textwidth]{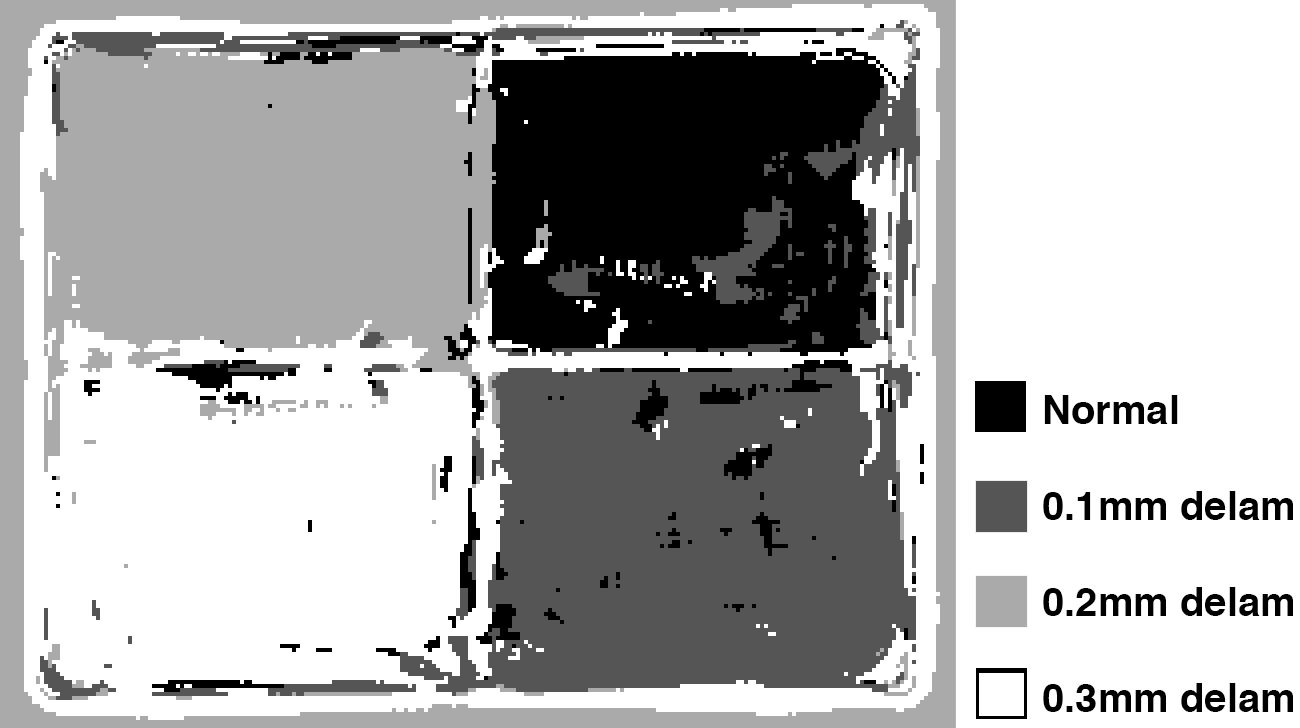}
 \caption{This figure shows the automatically-segmented image, representing each class as a different greyscale shade. The source information here is the noiseless set of features used in training and testing, so a portion of the dataset is insample. }
 \label{segmented}
\end{figure}

\begin{figure}[ht]
\centering
 \includegraphics[width=0.49\textwidth]{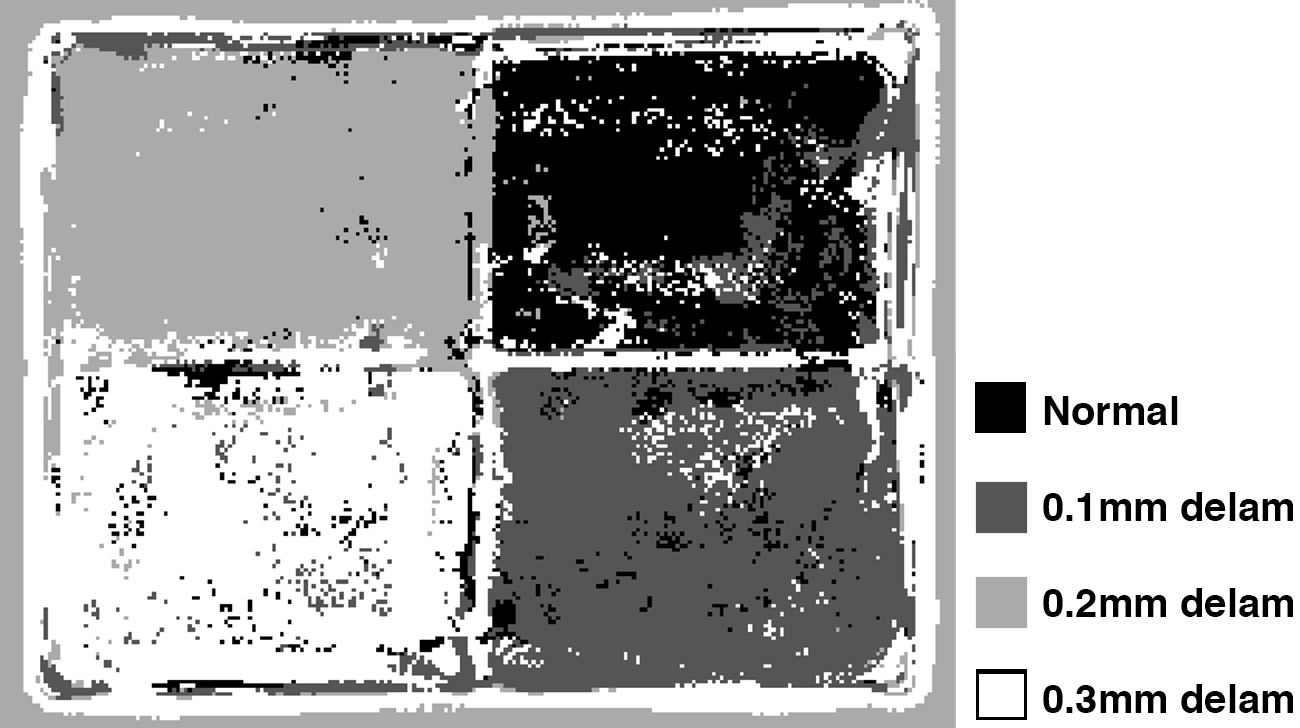}
 \caption{This figure shows the automatically-segmented image, representing each class as a different greyscale shade. The source information here is a set of modified per-pixel feature vectors with randomly applied noise up to $\pm3$\%. Note that while the classification accuracy is reduced, performance is still high when examining large regions.} 
 \label{segmented_noisy}
\end{figure}

The classification results for partial-insample are shown in Figure~\ref{segmented}, whereas the classification results for outsample data representing randomized $\pm3$\% noise addition is shown in Figure~\ref{segmented_noisy}. While insample performance is predictably better, outsample performance remains strong, and image-wise classification may be improved through the use of smoothing or clustering techniques. 

As the results show, the performance on real-world data reasonably matches that of the original DNN's performance for synthetic data. Interestingly, the TSR's noise-reducing capabilities appear to have a significant impact on model performance, as the outsample performance for real-world imaging does not differ widely from insample. This bodes well for model robustness and extensibility.

These results show that it is possible to automate non-destructive testing for common faults within FFF components manufactured from PLA. This will improve the technique's viability for unit-size orders for critical applications, enabling new business models for aerospace, healthcare, consumer, and other industries. 

\section{Conclusion and Future Work}\label{conclusion}
These results demonstrate that AI-TSR is a viable option to improve inspection rates for FFF components while reducing cost and time relative to manual inspection or capital-intensive approaches such as X-Ray imaging. This technique helps meet needs for high-throughput processes, allowing for $100$\% inspection rate and, when used in conjunction with an effective closed-loop feedback system, improving yields. The demonstrated technique is capable of identifying faults both due to incidental environmental effects, equipment wear, and tool failures, but also those faults maliciously executed as a result of compromised firmware, geometry files, or G-code. As a result, it may be possible to initiate equipment repairs, redesign processes, or to mitigate the damaging effects of a cyberattack more rapidly than is feasible for AM systems utilizing partial or low-automated inspection. Beyond increasing component strength and reliability, these results could one day improve resilience of multi-material components by ensuring complete intra- and inter-material adhesion or fusion. 

Continuing research into AI-TSR has the potential to enable bespoke manufacturing and late differentiation for AM (FFF or otherwise) in critical industries, while the use of features in excess of polynomial coefficients (for example MFCC, DWT, DFT, or kurtosis as used in \cite{misfire,afci}) offer avenues for accuracy improvement, and may provide sufficient granularity to extend the DNN to also measure delamination depth as well as thickness. In the future, we will print additional samples and extract TSR features to determine the limit of the classifier's sensitivity with regards to feature depth and delamination thickness, or to identify the minimum identifiable defect diameter and height in the case of defects such as voids. We intend to continue this work, extending inspection from FFF to other additive manufacturing techniques including vat photopolymerization, sheet lamination, directed energy deposition, binder jetting, and laser sintering. For each AM process, we hope to use an Instron to classify imaged sample properties to associate TSR features more closely with components' strength and likely failure mode. Additional features and samples will improve the classifier's gradation, leading to the development of continuous rather than discrete condition monitoring. 

The use of AI further facilitates full-automated inspection, e.g. as part of a production line. Features can be learned from known-good samples to provide a reference database for rapid detection of probable subsurface defects\cite{shepard2007automated}. The AI's output can then be used to request costly human inspection only in ambiguous cases, or to request targeted destructive testing to characterize a high probability fault and trace it back to specific points in the manufacturing process. Tied in to an IoT-enabled inventory management system, full inspection could allow component traceability and lead to improved manufacturer accountability, while the use of distributed recordkeeping with Blockchain-like systems could assure that component strength reports are simultaneously transparent and immutable. 

While the performance of automated non-destructive inspection and testing of FFF components using AI-TSR is compelling, there are opportunities for improvements such as the use of blob-detection to reduce sensitivity and improve robustness against noise and small-scale defects that may not be worthy of manual inspection, or alternative voting and/or segmentation techniques. Probable fault size or defect shape may be determined through pixel clustering, reporting the coordinates of sufficiently large faults to secondary human inspectors. Building this system into a mobile service\cite{tirecrack} or for embedded hardware\cite{afci} could further improve in-the-field utility, allows the classifier's use where conventional NDT imaging solutions are infeasible or where inspection latency and turnaround time is a critical concern. 

Finally, it remains to be seen how readily deployable this model will be for differing geometries. The trained DNN may be immediately portable, but more likely, will require the use of transfer learning to allow for customization for specific materials or geometries. Future work will validate the model's performance on true outsample data, as well as with other defect types, thin-walled geometries, and defects near edges in multiple common printing materials. Eventually, the DNN may be extended to support gradation, rather than classification -- for example to rank the degree of structural compromise within a sample. In this case, the classifier's activation sensitivity may be modified based on application type, e.g. with a $10$\% likelihood of failure raising an alert in aerospace, but a $50$\% probability used to trigger further inspection of consumer goods.

\section{Conflict of Interest}
Maria Beemer and Steven Shepard are employed by Thermal Wave Imaging, Inc. (TWI), which commercializes products utilizing TSR for component inspection. Josh Siegel approached TWI as a result of prior collaborations, and asked Maria and Steven to assist in imaging the FFF samples and generating TSR features as well as editing the NDT discussion within this manuscript. TWI volunteered its employees' time for this research, but made no financial contribution and was not involved with the creation of the neural network model or in computing the model's predictive accuracy. 

\section{Acknowledgements}
We gratefully acknowledge the support of NVIDIA Corporation for their donation of the Titan Xp graphics processing unit used to accelerate this research. 

\bibliographystyle{IEEEtran}
\bibliography{FDM}
\end{document}